\documentclass{ws-rv9x6}  

\renewcommand{\div}{\nabla\cdot}

\newcommand{\p}{\partial}
\newcommand{\EB}{\bm{E} \times \bm{B}}
\newcommand{\imag}{\mathrm{i}}
\newcommand{\diff}{\mathrm{d}}
\newcommand{\odf}[2]{\frac{\diff #1}{\diff #2}}
\newcommand{\pdf}[2]{\frac{\partial #1}{\partial #2}}

\usepackage{graphicx}	% required for `\includegraphics' (yatex added)
\begin{document}

\setcounter{chapter}{0}

\chapter{NONLINEAR SIMULATION OF DRIFT WAVE TURBULENCE}

\markboth{R. Numata}{Nonlinear Simulation of Drift Wave Turbulence}

\author{Ryusuke Numata, Rowena Ball, Robert L. Dewar}

\address{Department of Theoretical Physics, Research School of Physical
Sciences and Engineering,
The Australian National University, \\
Canberra ACT 0200, Australia \\
E-mail: ryusuke.numata@anu.edu.au}

\begin{abstract}
 In a two-dimensional version of the modified Hasegawa-Wakatani (HW)
 model, which describes electrostatic resistive drift wave turbulence,
 the  resistive coupling between vorticity and density does not act on
 the zonal components ($k_{y}=0$). It is therefore necessary to modify
 the HW model to treat the zonal components properly. The modified
 equations are solved numerically, and visualization and analysis of the
 solutions show generation of stable zonal flows, through conversion of
 turbulent kinetic energy, and the consequent turbulence and transport
 suppression. It is demonstrated by comparison that the modification is
 essential for generation of zonal flows. 
%The parameter that controls
%turbulence drive strength is used as a control parameter to construct
% simulated bifurcation diagrams.
\end{abstract}

\section{Introduction}     %S1-Heads

In quasi two-dimensional (2D) plasma and fluid flows the energy flux
from small scale turbulent modes toward lower wavenumber modes can 
dominate the classical Kolmogorov cascade to dissipative scales, with
the result that energy can accumulate in large scale coherent
structures. Zonal flows in planetary atmospheres and in magnetically
confined fusion plasmas are well-known examples of such coherent
structures. Quasi two-dimensional fluid systems in which turbulent
activities and coherent structures interact can undergo a spontaneous
transition to a turbulence-suppressed regime. In plasmas such
transitions dramatically enhance the confinement and are known as L--H or
confinement transitions. From theoretical and experimental works the
importance of shear or zonal flows for suppression of cross-field
transport and confinement improvement is now widely appreciated.

Several low-dimensional dynamical models, comprised of a small number of
  coupled ordinary differential equations, have been proposed to describe and predict the
  L--H transition\cite{diamond1994,sugamahorton1995,ball}. Ball {\it et al.}
  have analyzed a three-variable model using bifurcation and singularity
  theories\cite{ball}. The model is based on the reduced resistive
  magnetohydrodynamic equations with the electrostatic approximation,
  and describes the pressure-gradient-driven turbulence--shear flow
  energetics. This approach using low-dimensional modeling greatly
  simplifies the problem, and when validated against simulated or real
  experimental data, will provide an economical tool to predict
  transitions over the parameter space.

  In this work we report the results of numerical simulations that both
  complement the low-dimensional modeling results and raise some
  interesting issues in their own right. We focus on a model for
  electrostatic resistive drift wave turbulence, the Hasegawa-Wakatani
  (HW) model\cite{hw}, and solve the equations by direct numerical
  simulation in 2D slab geometry. The HW model has been
  widely used to investigate anomalous edge transport due to collisional
  drift waves\cite{driftstudy}. Moreover, self-organization of a shear
  flow   has been shown by numerical simulation of the HW model in
  cylindrical geometry\cite{hw_zonal}. Thus we consider the HW model is
  a good starting point for studying self-consistent turbulence--shear
  flow interactions, even though it does not describe physics that can
  be important in specific situations, such as magnetic curvature,
  magnetic shear, and electromagnetic effect.

\section{Modified Hasegawa-Wakatani Model}

The physical setting of the HW model may be considered as the edge
region of a tokamak plasma of nonuniform density $n_{0}=n_{0}(x)$ and in
a constant equilibrium magnetic field $\bm{B}=B_{0}\nabla z$. Following
the drift wave ordering\cite{hasegawa-mima}, the density $n=n_{0}+n_{1}$
and the electrostatic potential $\varphi$ perpendicular
to the magnetic field are governed by the continuity equation for ions
or electrons and the ion vorticity equation,
\begin{align}
 \odf{}{t} n & = \frac{1}{e} \pdf{}{z} j_{z},
  \label{eq:drifteq_n}\\
 \frac{mn}{B_{0}} \odf{}{t} \nabla_{\perp}^{2} \varphi & = B_{0}
  \pdf{}{z} j_{z},
  \label{eq:drifteq_v}
\end{align}
where $\nabla_{\perp}=\left(\p/\p x, \p/\p y\right)^{T}$,
$\diff/\diff t=\p/\p t+\bm{V}_{E}\cdot\nabla_{\perp}$ is the $\EB$
convective derivative ($\bm{V}_{E}\equiv-\nabla_{\perp} \varphi \times
\nabla z/B_{0}$, $\bm{E}=-\nabla_{\perp} \varphi$), $m$ is the 
ion mass, $j_{z}$ is the current density in the direction of the
magnetic field. The continuity equation (\ref{eq:drifteq_n}) can refer
to ions and electrons because $\div \bm{j}=0$ under the quasineutral
condition, and (\ref{eq:drifteq_v}) holds because the current density is
divergence-free. Since the ion inertia is negligible in the parallel
direction ($z$), the parallel current is determined by the Ohm's law,
\begin{equation}
 \bm{E} + \frac{1}{en} \nabla p_{\mathrm{e}} = \eta \bm{j}.
  \label{eq:ohm}
\end{equation}
If the parallel heat conductivity is sufficiently large, the electrons may
be treated as isothermal: $p_{\mathrm{e}}=nT_{\mathrm{e}}$
($p$ is the pressure, $T$ is the temperature, and subscript $\mathrm{e}$
refers to electrons.) This gives the parallel current as
\begin{equation}
 j_{z} = -\frac{1}{\eta} \pdf{}{z}
  \left(\varphi-\frac{T_{\mathrm{e}}}{e} \ln n \right). 
  \label{eq:jpara}
\end{equation}
If we eliminate $j_{z}$ from ($\ref{eq:drifteq_n}$),
($\ref{eq:drifteq_v}$) and normalize variables as
 \begin{equation}
  x/\rho_{\mathrm{s}} \rightarrow x, ~~
   \omega_{\mathrm{ci}} t \rightarrow t, ~~
   e \varphi / T_{\mathrm{e}} \rightarrow \varphi, ~~
   n_{1}/n_{0} \rightarrow n,
 \end{equation}
 where $\omega_{\mathrm{ci}}\equiv eB_{0}/m$ is the ion cyclotron
 frequency, and $\rho_{\mathrm{s}}\equiv
 \sqrt{T_{\mathrm{e}}/m}\omega_{\mathrm{ci}}^{-1}$ is the ion sound
 Larmor radius, we  finally obtain the resistive drift wave equations
 known as the Hasegawa-Wakatani (HW) model\cite{hw},
\begin{align}
 \pdf{}{t} \zeta + \{\varphi,\zeta \}
  & = \alpha(\varphi-n) - D_{\zeta} \nabla^{4}\zeta,
  \label{eq:hwv}\\
 \pdf{}{t} n + \{\varphi,n \}
 & = \alpha(\varphi-n) - \kappa \pdf{\varphi}{y} - D_{n}
 \nabla^{4} n,
 \label{eq:hwn}
\end{align}
where $\{a,b\}\equiv (\p a/\p x) (\p b/\p y) -(\p a /\p y) (\p b/\p x)$ is the
Poisson bracket, $\nabla^{2}=\p^{2}/\p x^{2}+\p^{2}/\p y^{2}$ is the
2D Laplacian, $\zeta\equiv\nabla^{2}\varphi$ is the vorticity. We omit
$\perp$, and use $\nabla$ for the 2D derivative. The dissipative terms
with constant  
coefficients $D_{\zeta}$ and $D_{n}$ have been included as adjuncts
without derivation, for numerical stability. The background density is
assumed to have an unchanging exponential profile: $\kappa \equiv
-(\p/\p x) \ln n_{0}$. $\alpha \equiv -T_{\mathrm{e}}/(\eta
n_{0}\omega_{\mathrm{ci}} e^{2}) \p^{2}/\p z^{2}$ is the adiabaticity
operator describing the parallel electron response. In a 2D setting the
coupling term operator $\alpha$ becomes a constant coefficient, or
parameter, by the replacement $\p/\p z \rightarrow \imag k_{z}$. This
resistive coupling term must be treated carefully in a 2D model because
zonal components of fluctuations (the $k_{y}=k_{z}=0$ modes) do not
contribute to the parallel current\cite{modification}. Recalling that
the tokamak edge turbulence is considered here, $k_{y}=0$ should always
coincide with $k_{z}=0$ because any potential fluctuation on the flux
surface is neutralized by parallel electron motion.
Let us define zonal and non-zonal components of a variable $f$ as
\begin{equation}
 \textrm{zonal:}~
  \langle f \rangle = \frac{1}{L_{y}}\int f \diff y,~~~
  \textrm{non-zonal:}~
  \tilde{f} = f - \langle f \rangle,
\end{equation}
where $L_{y}$ is the periodic length in $y$, and remove the contribution
by the zonal components in the resistive coupling term in (\ref{eq:hwv})
and (\ref{eq:hwn}). By subtracting the zonal components from the
resistive coupling term $\alpha (\varphi-n) \rightarrow
\alpha(\tilde{\varphi}-\tilde{n})$, we end up with the modified HW (MHW)
equations,
\begin{align}
 \pdf{}{t} \zeta + \{\varphi,\zeta \}
  & = \alpha(\tilde{\varphi}-\tilde{n}) - D_{\zeta} \nabla^{4}\zeta, 
  \label{eq:mhwv}\\
 \pdf{}{t} n + \{\varphi,n \}
 & = \alpha(\tilde{\varphi}-\tilde{n}) - \kappa \pdf{\varphi}{y} - D_{n}
 \nabla^{4} n.
 \label{eq:mhwn}
\end{align}
The evolution of the zonal components can be extracted from
(\ref{eq:mhwv}) and (\ref{eq:mhwn}) by averaging in the $y$ direction:
\begin{equation}
 \pdf{}{t} \langle f \rangle + \pdf{}{x}
  \left\langle f v_{x} \right\rangle
  = D \nabla^{2} \langle f
  \rangle,~~ v_{x}\equiv -\pdf{\tilde{\varphi}}{y},
  \label{eq:mhwz}
\end{equation}
where $f$ stands for $\zeta$ and $n$, and $D$ stands for the
corresponding dissipation coefficients.

The HW model spans two limits with respect to the adiabaticity
parameter. In the adiabatic limit $\alpha \rightarrow \infty$
(collisionless plasma), the non-zonal component of electron density obeys the
Boltzmann relation $\tilde{n}=n_{0}(x) \exp(\tilde{\varphi})$, and the
equations are reduced to the Hasegawa-Mima equation\cite{hasegawa-mima}. 
In the hydrodynamic limit $\alpha \rightarrow 0$ and the equations are
decoupled. Vorticity is determined by the 2D Navier-Stokes (NS)
equation, and the density fluctuation is passively advected by the flow
obtained from the NS equation.

In the ideal limit ($\alpha=\infty$, $D_{\zeta}=D_{n}=0$) the modified
HW system has two dynamical invariants, the energy $E$ and the
potential enstrophy $W$,
\begin{equation}
 E = \frac{1}{2} \int (n^{2}+|\nabla \varphi|^{2}) \diff\bm{x},~~~
  W = \frac{1}{2} \int (n-\zeta)^{2} \diff \bm{x},
\end{equation}
where $\diff\bm{x}=\diff x \diff y$, which constrain the fluid
 motion. According to Kraichnan's theory of 2D
 turbulence\cite{kraichnan1980}, the net flux of enstrophy is downscale
 while that of energy is upscale. This inverse energy cascade is behind
 the development of large scale, stable coherent structures in a HW flow.

 Conservation laws are given by
\begin{equation}
 \odf{E}{t} = \Gamma_{n} - \Gamma_{\alpha} - D_{E},~~~
  \odf{W}{t} = \Gamma_{n} - D_{W}.
\end{equation}
Fluxes and dissipations are given by
\begin{align}
 \Gamma_{n} & = - \kappa \int \tilde{n}\pdf{\tilde{\varphi}}{y} \diff\bm{x}, \\
 \Gamma_{\alpha} & =
% \alpha \int (\tilde{n}-\tilde{\varphi})(n-\varphi) \diff\bm{x} =
 \alpha \int (\tilde{n}-\tilde{\varphi})^{2} \diff\bm{x}, \\
 D_{E} & = \int [
  D_{n} (\nabla^{2} n)^{2} +
  D_{\zeta} |\nabla \zeta|^{2}
  ] \diff\bm{x},\\
 D_{W} & = \int [
  D_{n} (\nabla^{2} n)^{2} +
  D_{\zeta} (\nabla^{2} \zeta)^{2} -
  (D_{n}+D_{\zeta}) \nabla^{2} n \nabla^{2}\zeta
  ] \diff\bm{x}.
\end{align}
These quantities constitute sources and sinks. As will be seen in the
simulation results, they are mostly positive
($\Gamma_{\alpha}$ and $D_{E}$ are positive definite), thus only
$\Gamma_{n}$ can act as a source. The energy absorbed from the
background supplies the turbulent fluctuations through the drift wave
instability.

Note that the same conservation laws hold for the unmodified original HW
(OHW) model except that $\Gamma_{\alpha}$ is defined by both zonal and
non-zonal components; $\Gamma_{\alpha}^{\textrm{OHW}}\equiv
\alpha\int(n-\varphi)^{2}\diff \bm{x}$. In the OHW model, the zonal modes
as well as the non-zonal modes suffer the resistive dissipation.
%energy from the background through $\Gamma_{n}$, while it is only dissipated
%by $D$ in the MHW model.

\subsection{Linear Stability Analysis}

\begin{figure}
 \begin{center}
  \includegraphics[scale=0.7]{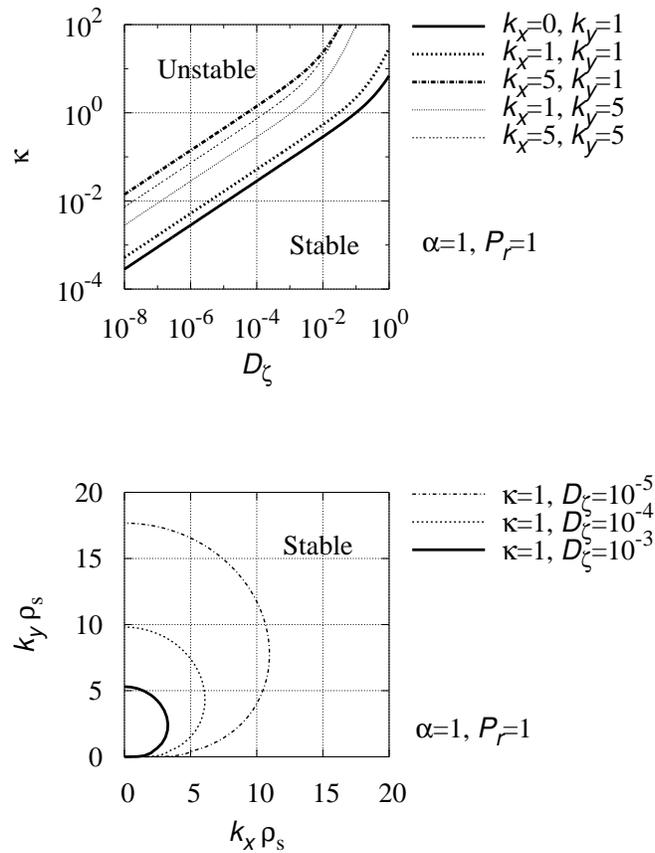}
  \caption{Stability diagram of the MHW model. Left panel shows the
  stability thresholds in $D_{\zeta}-\kappa$ plane. The drift wave
  instability can be stabilized by strong dissipation. In the right panel,
  stability thresholds are plotted in $k_{x}-k_{y}$ plane. For certain
  parameters, only some low wavenumber modes are unstable.}
  \label{fig:stabreg}
 \end{center}
\end{figure}

Since the zonal modes have linearly decaying solutions, we only consider
the form $e^{\imag (k_{x}x+k_{y}y-\omega t)}$
($k_{y}\neq0$). Linearization of the equations (\ref{eq:mhwv}) and
(\ref{eq:mhwn}) yields the dispersion relation, 
\begin{equation}
 \omega^{2} + \imag \omega (b+(1+P_{r}^{-1})k^{4} D_{\zeta}) -
  \imag b \omega_{\ast} - \alpha k^{2}(k^{2}+P_{r}^{-1})D_{\zeta} -
  k^{8} P_{r}^{-1} D_{\zeta}^{2} = 0,
  \label{eq:hw_dispersion}
\end{equation}
where we defined $k^{2}=k_{x}^{2}+k_{y}^{2}$, $b\equiv\alpha
(1+k^{2})/k^{2}$, the drift frequency $\omega_{\ast}\equiv k_{y}
\kappa/(1+k^{2})$, and the Prandtl number $P_{r} \equiv
D_{\zeta}/D_{n}$. Solutions to the dispersion relation
(\ref{eq:hw_dispersion}) are given by
\begin{align}
 \Re(\omega) & =
  \pm \frac{1}{2} (\sigma^{2}+16 b^{2} \omega_{\ast}^{2})^{\frac{1}{4}}
  \cos\frac{\theta}{2}, \\
 \Im(\omega) & =
  - \frac{1}{2}
  \left[
  b+(1+P_{r}^{-1})k^{4} D_{\zeta} \mp (\sigma^{2}+16
  b^{2}\omega_{\ast}^{2})^{\frac{1}{4}} \sin \frac{\theta}{2}
  \right],
\end{align}
$\sigma =
 4\alpha k^{2}(k^{2}+P_{r}^{-1})D_{\zeta}+
 4k^{8}P_{r}^{-1}D_{\zeta}^{2}-
 (b+(1+P_{r}^{-1})k^{4}D_{\zeta})^{2}$,
$\tan \theta = 4b\omega_{\ast}/\sigma$. In the limit where
 $D_{\zeta}=D_{n}=0$, it is readily proved that one of the growth rate 
 $\gamma\equiv \Im(\omega)$ is positive if $b\omega_{\ast}$ is finite,
 thus unstable. However, there exists a range of $D_{\zeta}$ where the
 drift wave instability is suppressed. The stability threshold is given
 by
 \begin{equation}
  \left(b+(1+P_{r}^{-1})k^{4}D_{\zeta}\right)^{4} \geq
   (\sigma^{2}+16b^{2}\omega_{\ast}^{2}) \sin^{4} \frac{\theta}{2},
 \end{equation}
 and is depicted in Fig.~\ref{fig:stabreg}. The left panel shows the
 stability boundary in $D_{\zeta}-\kappa$ plane. If we enhance the drive
 by increasing $\kappa$, the system becomes unstable. However, the
 instability is stabilized by increasing the dissipation. The stability
 threshold in $k_{x}-k_{y}$ plane is shown in the right panel. We see
 that in a highly driven-dissipative system only low wavenumber modes
 are unstable. The stability boundary in parameter space is a region
 where interesting dynamics are expected to occur, such as bifurcations
 or sudden changes to a suppressed (or enhanced) turbulence regime.

\begin{figure}
 \begin{center}
  \includegraphics[scale=0.45]{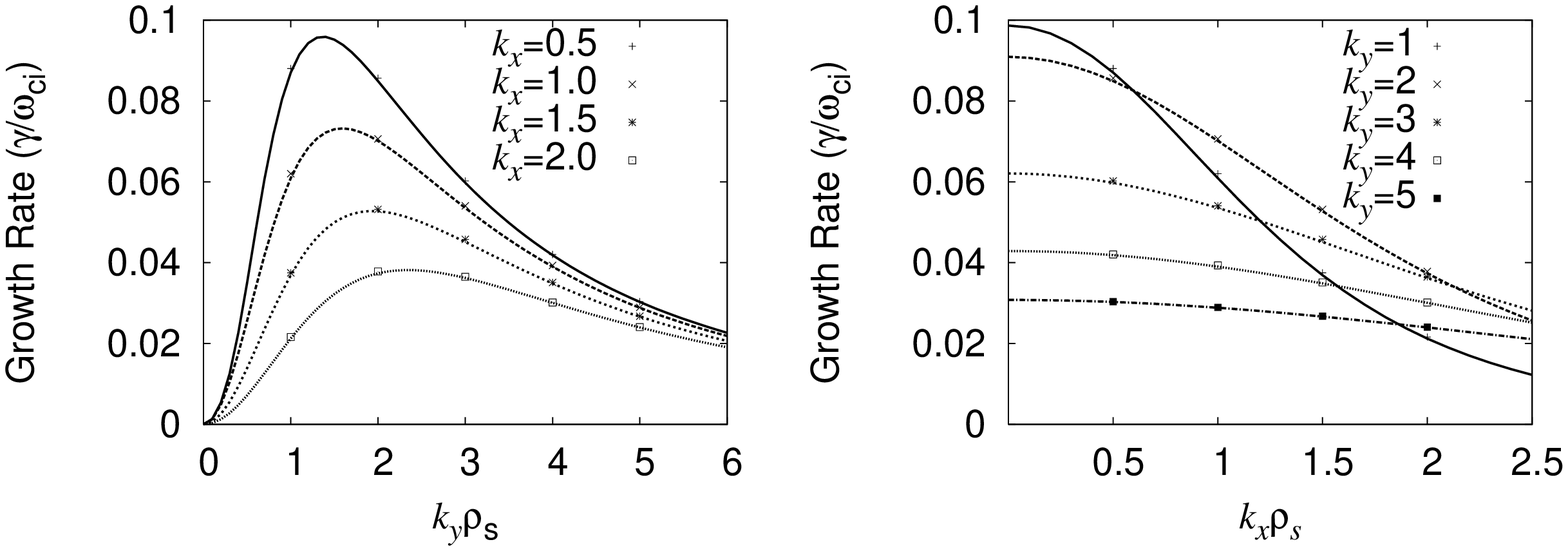}
  \caption{Dispersion relation of the dissipationless MHW
  model. $\alpha=1$, $\kappa=1$.}
  \label{fig:dispersion}
 \end{center}
\end{figure}

Figure~\ref{fig:dispersion} shows the dispersion relation for
cases where $D_{\zeta}=D_{n}=0$. To provide a test of the simulation
code, we plot growth rates obtained from numerical simulations together
with the analytic curves. We can see that the growth rates obtained
numerically agree very well with that calculated analytically. We also
note that, in the parameter range plotted in Fig.~\ref{fig:dispersion}
($\alpha=1$, $\kappa=1$), the most unstable mode is $k_{x}\sim0$,
$k_{y}\sim 1$.

\section{Simulation Results}

The HW equations are solved in a double periodic slab domain with box
size $(2L)^{2}=(2\pi/\Delta k)^{2}$ where the lowest wavenumber
$\Delta k=0.15$. The equations are discretized on $256\times256$ grid
points by the finite difference method. Arakawa's method is used for
evaluation of the Poisson bracket\cite{arakawa}. Time stepping algorithm
is the third order explicit linear multistep method\cite{bdf}.

Since we are focusing in this work on how the modification
(\ref{eq:mhwv}), (\ref{eq:mhwn}) influences nonlinearly saturated
states, we fix the parameters to $\kappa=1$, $D_{\zeta}=10^{-6}$,
$\alpha=1$, and $P_{r}=1$, and compare the results obtained using the
MHW model with those computed from the OHW model. For these parameters
the system is unstable for most wavenumbers. During a typical evolution,
initial small amplitude perturbations grow linearly until the nonlinear
terms begin to dominate. Then the system arrives at a nonlinearly
saturated state where the energy input $\Gamma_{n}$ and output due to
the resistivity $\Gamma_{\alpha}$ and the dissipations $D_{E,W}$ balance.

\begin{figure}[htbp]
 \begin{center}
  \includegraphics[scale=0.7]{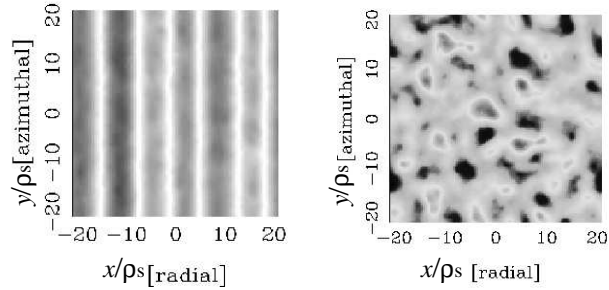}
  \caption{Contour plots of saturated electrostatic potentials for
  the modified and the original HW models. Zonally elongated structure
  is clearly visible for MHW case.}
  \label{fig:zonal}
 \end{center}
\end{figure}

\begin{figure}[htbp]
 \begin{center}
  \includegraphics[scale=0.5]{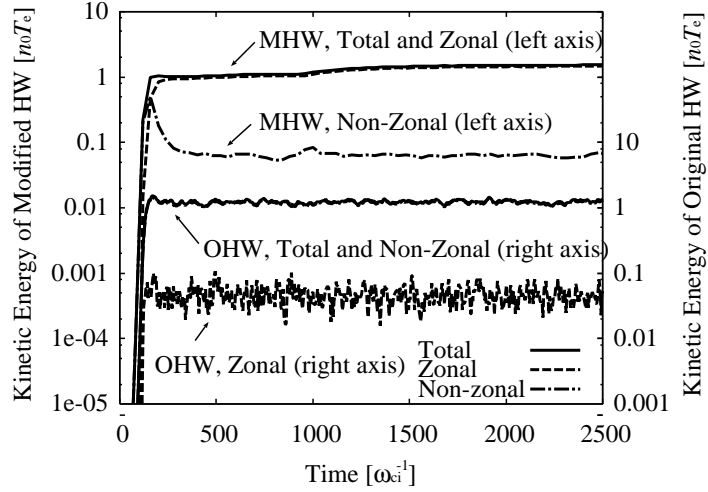}
  \caption{Time evolution of the kinetic energy, and its partition to
  the zonal and the non-zonal components. In the modified HW model, the
  zonal mode contains most of kinetic energy, while non-zonal turbulence
  contains most of the kinetic energy in the original HW model.}
  \label{fig:kinetic}
 \end{center}
\end{figure}

In Fig.~\ref{fig:zonal}, we contrast the zonally elongated structure of
the saturated electrostatic potential computed from the MHW model with
the strong isotropic vortices in that from the OHW model. Time evolution
of the kinetic energy $E^{\mathrm{K}}=1/2\int|\nabla\varphi|^{2}\diff
\bm{x}$, and its partition to the zonal and the non-zonal components are
shown in Fig.~\ref{fig:kinetic}. The saturated kinetic energy is not
affected by the modification ($E^{\mathrm{K}}\sim 1$ for both cases). In
the OHW model, the zonal flow grows in  the linear phase, as well as the
other modes, up to a few percent of the kinetic energy, and
saturates. On the other hand, in the MHW model the zonal kinetic energy
continues to grow after the linear phase, and dominates the kinetic
energy. The kinetic energy contained in other modes decreases to a few
percent of the total kinetic energy. In the original 2D HW model, the
resistive coupling term is retained for the zonal modes, the effect of
which is to prevent development of zonal flows. But since the zonal
modes do not carry parallel currents it is clearly unphysical to retain
resistive action on them. Subtraction of the zonal components from the
resistive coupling term is necessary to permit the generation of zonal
flows.

\begin{figure}[htbp]
 \begin{center}
  \includegraphics[scale=0.5]{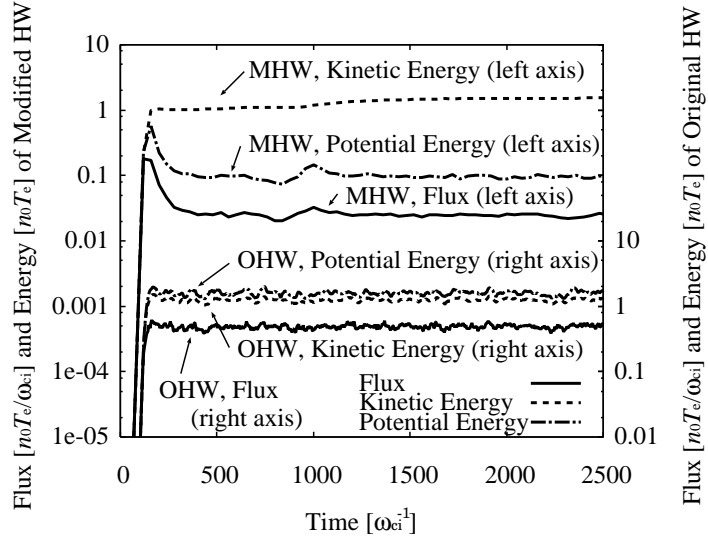}
  \caption{Time evolutions of the radial density transport and the
  kinetic and the potential energies for the modified and the original HW models.
  Once zonal flow is generated in MHW model, the turbulent fluctuation
  level and transport are significantly reduced.}
  \label{fig:transport}
 \end{center}
\end{figure}

The density flux in $x$ direction $\Gamma_{n}$ (transport across the
magnetic field), together with the energy partition to the kinetic energy
$E^{\mathrm{K}}$ and the potential energy $E^{\mathrm{P}}=1/2\int
n^{2}\diff \bm{x}$, is plotted in Fig.~\ref{fig:transport}. We observe
that once the zonal flow is generated in the MHW model, the transport
level is significantly suppressed. The transport suppression is mostly
because the saturated potential energy (or amplitude of saturated density
fluctuation) is reduced. The potential energy and the turbulence kinetic
energy are converted into the zonal kinetic energy. By contrast the
energy of the OHW model is almost equi-partitioned between the kinetic and
potential energy.

\begin{figure}[htbp]
 \begin{center}
  \includegraphics[scale=0.7]{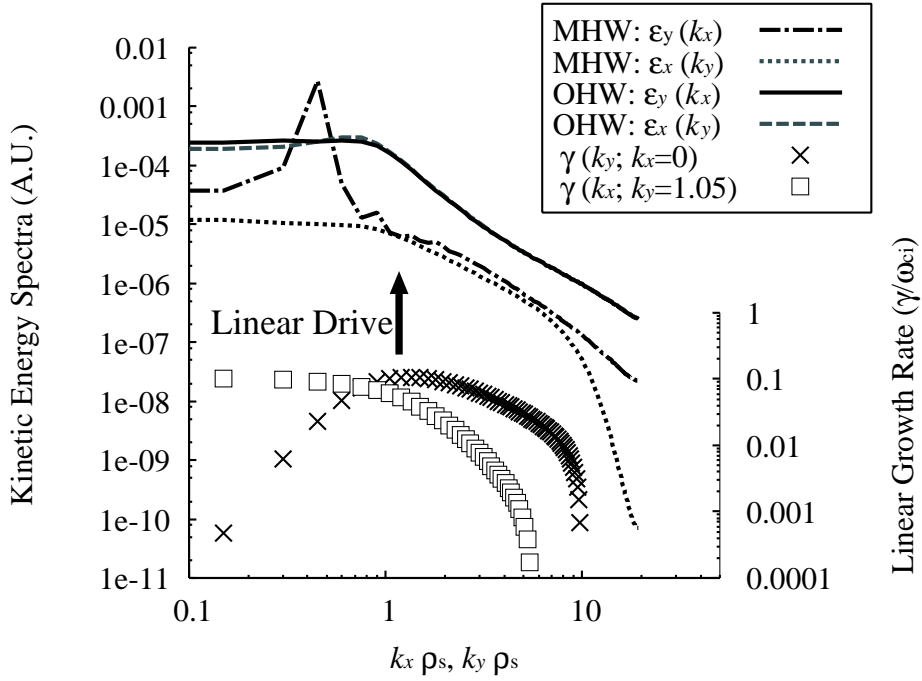}
  \caption{The $x$ and $y$ averaged kinetic energy spectra for the MHW
  and the OHW models. The top two lines (solid line for
  ${\mathcal E}_{y}(k_{x})$ and broken line for
  ${\mathcal E}_{x}(k_{y})$) for the OHW model are almost
  overlapped indicating isotropy. The middle two lines (dot-dashed
  line for ${\mathcal E}_{y}(k_{x})$ and dotted line for ${\mathcal
  E}_{x}(k_{y})$) for MHW show highly anisotropic structure in low
  $k$ region. The energy injected at $(k_{x}, k_{y})=(0,1)$ cascades
  inversely to the zonal mode of the wave number $(0.45,0)$. The bottom
  two series of symbols show the linear growth rates of modes for reference.}
  \label{fig:spectra}
 \end{center}
 \end{figure}

The kinetic energy spectra averaged over the $x$ or $y$ direction for the
MHW and the OHW models are shown in Fig.~\ref{fig:spectra}. The $x$ ($y$)
averaged kinetic energy spectra (${\mathcal E}_{x (y)}^{\mathrm{K}}$)
are defined from the Fourier amplitude of the kinetic energy ${\mathcal
E}^{\mathrm{K}}$ by  
\begin{align}
 {\mathcal E}_{y}^{\mathrm{K}}(k_{x}) & = \frac{1}{K_{y}}\int_{0}^{K_{y}}
  {\mathcal E}^{\mathrm{K}}(k_{x},k_{y}) \diff k_{y}, \\
 {\mathcal E}_{x}^{\mathrm{K}}(k_{y}) & = \frac{1}{K_{x}}\int_{0}^{K_{x}}
  {\mathcal E}^{\mathrm{K}}(k_{x},k_{y}) \diff k_{x},
\end{align}
where $K_{x}, K_{y}$ are the highest wavenumbers. The spectra of the
modified model again show strong anisotropic structure whereas there
is no marked difference in the original HW model. In the modified model,
potential energy stored in the background density is converted into
turbulent kinetic energy through the drift wave instability at $k_{y} \sim
1$, $k_x=0$ and then is distributed to smaller wavenumbers. The drift wave
structure, which is elongated in the $x$ direction, is break up into
rather isotropic vortices after the nonlinear effect sets in, and those
isotropic vortices merge in the $y$ direction to produce the zonal
flow. We can recognize this non-negligible inverse energy cascade in the
$y$ direction from a slight negative slope of ${\mathcal E}_{x}(k_{y})$
spectrum in $k_{y} \lesssim 1$ region. The $y$ averaged spectrum
${\mathcal E}_{x}(k_{y})$ shows the strong peak at the zonal wave number
$k_{x}\sim 0.45$.

\section{Conclusion}

We have performed nonlinear simulations of the 2D HW model. As suggested
recently\cite{modification}, the electron response parallel to the
background magnetic field must be treated carefully in the 2D model. The
model should be modified to exclude the zonal ($k_{y}=0$) contribution
from the resistive coupling term. By comparing the numerical results of
the modified and the unmodified original HW models, we have revealed
that a remarkable zonal flow structure in the nonlinearly saturated
state is only observed in the modified model. Thus, the modification is
crucial to the generation of the zonal flow in this model. Time
evolutions of the macroscopic quantities, such as the energies and
fluxes show that, after the zonal flow is built up by turbulent
interaction, the generated zonal flow significantly suppresses the
turbulent fluctuation level and the cross-field density transport.

The build up of the zonal flow and resulting transport suppression
indicate bifurcation structure of the system. If we increase a parameter
(say, strength of the linear drive term $\kappa$), the system may
undergo sudden transition from a high transport to a low transport regime.
The state shown in this paper can be a bifurcated state. A systematic
parameter study and comparison with the low-dimensional dynamical model
are possible next steps.

\section*{Acknowledgments}

The simulation code used in this paper is provided by B.D. Scott.
The authors would like to thank J.A. Krommes, F. Jenko and H.A. Dijkstra
for fruitful discussions and comments during the Workshop on Turbulence
and Coherent Structures. This work is supported by the Australian
Research Council.

\end{document}